\documentclass[12pt]{article}
\usepackage{epsfig}
\usepackage{amssymb}
\usepackage{amsmath}
\usepackage{amsfonts}

\newcommand{\R}{\mathbb{R}}

\newcommand{\be}{\begin{equation}}
\newcommand{\ee}{\end{equation}}
\newcommand{\bea}{\begin{eqnarray}}
\newcommand{\eea}{\end{eqnarray}}
\newcommand{\nn}{\nonumber}
\newcommand{\kt}{\rangle}
\newcommand{\br}{\langle}

\newcommand{\cum}{\mbox{\scriptsize${\cal M}$}}
\newcommand{\ed}{\end{document}}

\newcommand{\rx}{{\rm x}}

\newcommand{\rp}{{\rm p}}

\newcommand{\rH}{{\rm H}}
\newcommand{\rX}{{\rm X}}
\newcommand{\rP}{{\rm P}}
\newcommand{\rh}{{\rm h}}
\newcommand{\rE}{{\rm E}}

\newcommand{\np}{\newpage}

\begin{document}

\title{${\cal PT}$-Symmetric Cubic Anharmonic
Oscillator\\ as a Physical Model}

\author{\\
Ali Mostafazadeh
\\
\\
Department of Mathematics, Ko\c{c} University,\\
34450 Sariyer, Istanbul, Turkey\\ amostafazadeh@ku.edu.tr}
\date{ }
\maketitle

\begin{abstract}
We perform a perturbative calculation of the physical observables,
in particular pseudo-Hermitian position and momentum operators,
the equivalent Hermitian Hamiltonian operator, and the classical
Hamiltonian for the ${\cal PT}$-symmetric cubic anharmonic
oscillator, $H=\frac{1}{2m}p^2+\frac{1}{2}\mu^2x^2+i\epsilon x^3$.
Ignoring terms of order $\epsilon^4$ and higher, we show that this
system describes an ordinary quartic anharmonic oscillator with a
position-dependent mass and real and positive coupling constants.
This observation elucidates the classical origin of the reality
and positivity of the energy spectrum. We also discuss the
quantum-classical correspondence for this ${\cal PT}$-symmetric
system, compute the associated conserved probability density, and
comment on the issue of factor-ordering in the pseudo-Hermitian
canonical quantization of the underlying classical system.
\end{abstract}
\begin{center}
~~~~~PACS numbers: 03.65.-w
\end{center}
\vspace{2mm}




\np

\section{Introduction}

${\cal PT}$-symmetric quantum mechanics was originated by the
observation, initially made by Bessis and Zinn-Justin, that the
Hamiltonian for a cubic anharmonic oscillator:
    \be
    H=\frac{p^2}{2m}+\frac{\mu^2}{2}\,x^2+i\epsilon\, x^3,
    \label{H}
    \ee
with $\mu,\epsilon\in\R$ has a real, positive, and discrete
spectrum. During the past six years there have appeared a number
of publications \cite{bender-98}-\cite{bender-prd-2004} exploring
the properties of the Hamiltonian (\ref{H}). Yet the nature of the
physical system described by this Hamiltonian has not been
clarified. The present article aims at addressing this basic
issue. We will achieve this aim by computing the physical
observables, the localization probability density, and the
underlying classical Hamiltonian for this system. This is the
first example of a ${\cal PT}$-symmetric quantum system with
configuration space $\R$ that allows for such a computation.

As the main technical tools used in our analysis have been
developed in the study of the spectral properties of ${\cal
PT}$-symmetric Hamiltonians, we will include here a brief account
of the relevant developments.

The first convincing numerical evidence supporting the reality and
positivity of the spectrum of (\ref{H}) was provided by Bender and
Boettcher \cite{bender-98} who made the important observation that
this Hamiltonian was ${\cal PT}$-symmetric. Among dozens of
publications on the subject that followed \cite{bender-98} was the
article by Dorey, Dunning, and Tateo \cite{dorey-2001} that
provided the first mathematically rigorous proof of the spectral
properties conjectured by Bessis and Zinn-Justin (See also
\cite{shin}). From a physicist's point of view, a more important
development was the idea, put forward by Bender and his
collaborators \cite{bender-98,bender-99}, that such ${\cal
PT}$-symmetric Hamiltonians might be used as the Hamiltonian
operator for an extended/generalized quantum theory.

The main obstacle for realizing this idea was that a non-Hermitian
Hamiltonian such as (\ref{H}) generated a nonunitary
time-evolution. This was not compatible with the conventional
probabilistic interpretation of quantum mechanics. The resolution
of this problem came as a by-product of the attempts to
characterize the non-Hermitian operators having a real spectrum,
\cite{p1} -- \cite{jmp-2003}.

Ref.~\cite{p2} lists the necessary and sufficient conditions that
ensure the reality of the spectrum of a diagonalizable
operator.\footnote{In view of the requirements of the standard
quantum measurement theory physical observables in general and the
Hamiltonian in particular must be diagonalizable operators
\cite{p61}.} Among these is the condition that $H$ must be
Hermitian with respect to a positive-definite inner product
$\br\cdot,\cdot\kt_+$. This inner product is generally different
from the defining inner product $\br\cdot,\cdot\kt$ of the
(reference) Hilbert space ${\cal H}$ in which the operator $H$
acts. It can be conveniently expressed in terms of a
positive-definite (metric) operator $\eta_+:{\cal H}\to{\cal H}$
according to \cite{p1}
    \be
    \br\cdot,\cdot\kt_+=\br\cdot,\eta_+\cdot\kt.
    \label{inn}
    \ee
The condition that $H$ be Hermitian with respect to
$\br\cdot,\cdot\kt_+$, i.e., $\br\cdot,H\cdot\kt_+=\br
H\cdot,\cdot\kt_+$, is equivalent to $\eta_+$-pseudo-Hermiticity
of $H$, \cite{p1}. This means that $\eta_+$ belongs to the set
${\cal U}_H$ of all Hermitian invertible operators $\eta:{\cal
H}\to{\cal H}$ satisfying\footnote{Here and throughout this paper
the adjoint of an operator $A:{\cal H}_1\to{\cal H}_2$ between two
Hilbert spaces ${\cal H}_1$ (with inner product
$\br\cdot,\cdot\kt_1$) and ${\cal H}_2$ (with inner product
$\br\cdot,\cdot\kt_2$) is defined to be the unique operator
$A^\dagger:{\cal H}_2\to{\cal H}_1$ satisfying
$\br\cdot,A\cdot\kt_2=\br A^\dagger\cdot,\cdot\kt_1$.} \cite{p1}
    \be
    H^\dagger=\eta H\eta^{-1}.
    \label{ph}
    \ee
An interesting property of the set ${\cal U}_H$ of all metric
operators $\eta$ is that to each pair $(\eta_1,\eta_2)$ of
elements of ${\cal U}_H$ there corresponds a symmetry generator
$\eta_2^{-1}\eta_1$ of $H$, \cite{p1}. Furthermore ${\cal
PT}$-symmetric Hamiltonians $H$ that act in ${\cal H}=L^2(\R)$,
e.g., (\ref{H}), are ${\cal P}$-pseudo-Hermitian, i.e., ${\cal
P}\in{\cal U}_H$. This in turn implies that if $H$ has a real
spectrum then ${\cal P}^{-1}\eta_+={\cal P}\eta_+$ commutes with
$H$. The construction of the physical Hilbert space ${\cal H}_{\rm
phys}$ that is based on the ${\cal CPT}$-inner product \cite{bbj}
makes an implicit use of this observation. As shown in
\cite{jmp-2003} for theories defined on $\R$ and more recently
generalized in \cite{p62} to theories defined on a complex
contour, the ${\cal C}$ operator introduced in \cite{bbj} is
related to the metric operator $\eta_+$ according to
    \be
    {\cal C}={\cal P}\eta_+=\eta_+^{-1}{\cal P},
    \label{C}
    \ee
and the ${\cal CPT}$-inner product is precisely
$\br\cdot,\cdot\kt_+$.

The recent approximate calculations of ${\cal C}$ for the
anharmonic oscillator (\ref{H}) and its analogs
\cite{bender-prd-2004,bender-pla-2004} have also revealed the
practical significance of the factorization (\ref{C}) of ${\cal
C}$. These calculations are based on Eq.~(\ref{C}) and the
observation that (being a positive-definite operator) $\eta_+$
admits an exponential representation,
    \be
    \eta_+=e^{-Q}
    \label{eta=}
    \ee
where $Q$ is a Hermitian operator.

The metric operator $\eta_+$ (which is generally unique up to
symmetries of $H$, \cite{p4,jmp-2003,jmp-2004}), not only
determines the structure of the physical Hilbert space but it
fixes the observables of the theory as well,
\cite{critique,cjp-2004b,p61}. By definition ({\bf Def.~1.}) {\em
Physical observables are the Hermitian operators acting in the
physical Hilbert space ${\cal H}_{\rm phys}$},
\cite{critique,cjp-2004b,p61}, i.e., $A:{\cal H}\to{\cal H}$ is an
observables if\footnote{As argued in \cite{critique} identifying
observables with ${\cal CPT}$-invariant operators as initially
done in \cite{bbj} leads to a dynamical inconsistency. The latter
is avoided if one modifies this definition as proposed in
\cite{bender-erratum}. This modified definition is equivalent to
Def.~1 above for symmetric Hamiltonians $H$ (satisfying $\br x| H
|x'\kt=\br x'| H |x\kt$) and cannot be applied for nonsymmetric
Hamiltonians \cite{comment}.}
    \be
    \br\cdot,A\cdot\kt_+=\br A\cdot,\cdot\kt_+.
    \label{def}
    \ee
Alternatively, {\em Physical observables $A$ are
$\eta_+$-pseudo-Hermitian operators acting in ${\cal H}$}, i.e.,
$A^\dagger=\eta_+A\,\eta_+^{-1}$.

In order to see the central role played by the metric operator
$\eta_+$ in the construction of the observables, we recall that as
an operator mapping ${\cal H}_+$ onto ${\cal H}$ the unique
positive square root $\rho=\sqrt\eta_+$ of $\eta_+$ is a unitary
operator \cite{jpa-2003,p61}, i.e.,
    \be
    \br\rho\;\cdot,\rho\;\cdot\kt=\br\cdot,\cdot\kt_+.
    \label{unitary}
    \ee
Hence the Hermitian operators $O$ acting in ${\cal H}_{\rm phys}$
(i.e., the physical observables) may be obtained from the
Hermitian operators $o$ acting in ${\cal H}$ according to
    \be
    O=\rho^{-1}\,o\,\rho.
    \label{obs}
    \ee
This is also consistent with the condition \cite{p2} that $H$ is
related to a Hermitian operator $h:{\cal H}\to{\cal H}$ by a
similarity transformation,
    \be
    h=\rho\, H\,\rho^{-1}.
    \label{h=}
    \ee
The mapping $\rho:{\cal H}_{\rm phys}\to{\cal H}$ establishes the
unitary equivalence of the ${\cal PT}$-symmetric quantum system
${\cal S}_{PT}$ having ${\cal H}_{\rm phys}$, $H$, and $O$ as the
physical Hilbert space, the Hamiltonian, and the physical
observables and the quantum system ${\cal S}$ having ${\cal H}$,
$h$, and $o$ as the physical Hilbert space, the Hamiltonian, and
the physical observables, respectively, \cite{critique,p61}.
${\cal S}_{PT}$ and ${\cal S}$ describe the same physical system
because the physical quantities such as the expectation values and
transition amplitudes are independent of the choice of ${\cal
S}_{PT}$ and ${\cal S}$.

The advantage of the ${\cal PT}$-symmetric description provided by
${\cal S}_{PT}$ over the Hermitian description provided by ${\cal
S}$ is that unlike $H$, the Hermitian Hamiltonian $h$ is generally
nonlocal. This advantage is however balanced by the disadvantage
that the physical (pseudo-Hermitian) position $X$ and momentum
operators $P$ of ${\cal S}_{PT}$ are also generally nonlocal.
These operators are defined by \cite{p62}
    \be
    X:=\rho^{-1}\,x\,\rho,~~~~~~~~P:=\rho^{-1}\,p\,\rho,
    \label{X-P}
    \ee
where $x$ and $p$ are the conventional position and momentum
operators. The main advantage of the Hermitian description is that
it provides means for identifying the underlying classical system,
\cite{p61}. The classical Hamiltonian is obtained by expressing
$h$ in terms of $x$ and $p$ and replacing the latter with the
classical (real-valued) position $x_c$ and momentum $p_c$
observables. In general this yields an expression that may involve
powers of $\hbar$. The classical Hamiltonian $H_c$ is then
obtained by evaluating this expression in the limit $\hbar\to 0$,
i.e., assuming that this limit exists,
    \be
    H_c(x_c,p_c):=\lim_{\hbar\to 0} h(x_c,p_c).
    \label{class}
    \ee
The initial Hamiltonian $H$ may be recovered by performing the
so-called $\eta_+$-pseudo-Hermitian canonical quantization of
$H_c$ and adopting an appropriate factor-ordering prescription
\cite{p61}. Disregarding the complications due to the
factor-ordering problem and assuming that $H_c$ is an analytic
function of $x_c$ and $p_c$, we have
    \be
    H_c(X,P)=h(X,P)=h(\rho^{-1}\,x\,\rho,\rho^{-1}\,p\,\rho)=
    \rho^{-1}\,h(x,p)\,\rho=H.
    \label{H-class}
    \ee

Having introduced the $\eta_+$-pseudo-Hermitian position operator
$X$ we can also address the issue of determining the conserved
probability density $\rho$ for the localization of the system in
the configuration space. This requires the identification of the
physical localized states of the system. Being (the generalized
\cite{bohm-qm}) eigenvectors of $X$, the localized state vectors
are given by
    \be
    |\xi^{(x)}\kt=\rho^{-1}|x\kt,
    \label{loc}
    \ee
where $|x\kt$ are the conventional position eigenvectors. The
conserved probability density associated with a given state vector
$\psi\in{\cal H}_{\rm phys}$ has the form \cite{p61}
    \be
    \varrho(x)=N^{-1}|\Psi(x)|^2,
    \label{rho}
    \ee
where $\Psi(x)$ is the physical position wave function for the
state vector $\psi$, i.e.,
    \be
    \Psi(x):=\br \xi^{(x)},\psi\kt_+=\br x|\rho\,\psi\kt,
    \label{wf}
    \ee
$\br\cdot|\cdot\kt$ is the usual $L^2$-inner product on ${\cal
H}=L^2(\R)$, and $N:=\br\psi,\psi\kt_+=\int_{-\infty}^\infty
|\Psi(x)|^2dx$.\footnote{The physical position wave functions
evolve in time according to the Schr\"odinger equation with $h$
being the Hamiltonian operator. In general, one can express $h$ in
the form $p^2/(2m)+W$ where $W$ is a nonlocal potential (an
infinite series in $p$ with $x$-dependent coefficients). This in
turn implies that the probability current density that together
with $\varrho$ satisfies the continuity equation has a nonlocal
dependence on $\Psi(x)$; it is not given by the standard formula,
unless $H$ is Hermitian. This is especially significant in the
study of tunnelling and scattering for pseudo-Hermitian
Hamiltonians (having scattering states).}

An important feature of the exponential representation
(\ref{eta=}) of the metric operator $\eta_+$ is that it reduces
the calculation of $\rho$ and $\rho^{-1}$ to that of $Q$, for
    \be
    \rho^{\pm 1}=e^{\mp Q/2}.
    \label{rho=}
    \ee
We will use this observation together with the approach pursued in
\cite{bender-prd-2004} to perform a perturbative calculation of
$X$, $P$, $h$, $H_c$, and $\varrho$ for the ${\cal PT}$-symmetric
Hamiltonian~(\ref{H}).

Because we are interested in the issue of finding the classical
limit of the ${\cal PT}$-symmetric theory based on the Hamiltonian
(\ref{H}) we wish to retain the factors of $\hbar$. However, for
the simplicity of the calculations and ease of the comparison with
the known results, we will introduce and employ the following
dimensionless quantities.
    \bea
    \rx&:=&\ell^{-1}x,~~~~~~~~~~~~~~~~~~~~~~
    ~~~~~~~~~\rp:=\ell \hbar^{-1}p,
    \label{new1}\\
    \cum &:=&\ell^2\hbar^{-1}\sqrt m\, \mu,~~~~~~~~~~~~~~~~~~~~~~
    \varepsilon:=\ell^5\hbar^{-2}m\,\epsilon,
    \label{new2}\\
    \rH_0&:=&\frac{1}{2}\,\rp^2+
    \frac{\cum^2}{2}\,\rx^2,~~~~~~~~~~~~~~~~~~~~
    \rH_1:=i\,\rx^3,
    \label{new3}\\
    \rH&:=&\ell^2\hbar^{-2}m\, H=
    \frac{1}{2}\,\rp^2+\frac{\cum^2}{2}\,\rx^2+i\,\varepsilon\,
    \rx^3=\rH_0+\varepsilon\,\rH_1,
    \label{rH}
    \eea
where $\ell$ is an arbitrary length scale which may be taken as
$\mu^2/\epsilon$. Clearly, we have $[\rx,\rp]=i$.

\section{Calculation of $Q$}

In \cite{bender-prd-2004} the authors outline a perturbative
calculation of $Q$ for the Hamiltonian (\ref{rH}) taking
$\varepsilon$ as the perturbation parameter. They use the
identities $[{\cal C},{\cal PT}]=0$ and ${\cal C}^2=1$ to infer
that as a function of $\rx$ and $\rp$, $Q$ must be even in $\rx$
and odd in $\rp$. Furthermore, imposing $[C,\rH]=0$ and making use
of the fact that $\rH_1$ is an imaginary cubic potential, they
find the operator equation
    \be
    2\,\varepsilon\, e^Q \rH_1=[e^Q,\rH],
    \label{condi-1}
    \ee
and that $Q$ may be expanded in an odd power series in
$\varepsilon$,
    \be
    Q=Q_1\varepsilon+Q_3\varepsilon^3+Q_5\varepsilon^5+\cdots,
    \label{Q=}
    \ee
where $Q_{2i+1}=Q_{2i+1}(x,p)$ with $i=0,1,2,\cdots$ are
$\varepsilon$-independent. Next, they expand $e^Q$ in power series
in $\varepsilon$, substitute the result in (\ref{condi-1}), and
demand that this equation be satisfied at each order of
$\varepsilon$. This yields a series of operator equations that
they iteratively solve for $Q_{2i+1}$.

The operator equations whose solution yield $Q_{2i+1}$ may be more
conveniently obtained from the $\eta_+$-pseudo-Hermiticity of
$\rH$,
    \be
    \rH^\dagger=\eta_+\,\rH\,\eta_+^{-1}.
    \label{ph+}
    \ee
Substituting $\eta_+=e^{-Q}$ in this equation and noting that
$\rH^\dagger=\rH_0-\epsilon\, \rH_1$, we have
    \be
    \rH_0-e^{-Q}\rH_0\, e^Q=\epsilon(\rH_1+e^{-Q}\rH_1\, e^Q).
    \label{c1}
    \ee
Next, we employ the Baker-Campbell-Hausdorff identity,
    \be
    e^{-A}B\,e^A= B+[B,A]+\frac{1}{2!}[[B,A],A]+\frac{1}{3!}
    [[[B,A],A],A]+\cdots,
    \label{bch}
    \ee
(where $A$ and $B$ are linear operators), to express (\ref{c1}) as
{\small
    \bea
    -[\rH_0,Q]-\frac{1}{2!}\,[[\rH_0,Q],Q]-
    \frac{1}{3!}\,[[[\rH_0,Q],Q],Q]-\cdots&=&\varepsilon\left(
    2\rH_1+[H_1,Q]+
    \frac{1}{2!}\,[[\rH_1,Q],Q]+\right.\nn\\
    &&\quad\quad\quad\left.
    \frac{1}{3!}\,[[[\rH_1,Q],Q],Q]+\cdots\right).
    \label{c2}
    \eea}
\noindent Now, in view of (\ref{Q=}), we can easily identify the
terms in (\ref{c2}) that are of the same order in powers of
$\varepsilon$. Enforcing (\ref{c2}) at each order, we find the
desired operator equations for $Q_{2i+1}$. Matching the terms of
order $\varepsilon,\varepsilon^2,\cdots,\varepsilon^5$, we find in
this way the following independent operator equations which agree
with those obtained in \cite{bender-prd-2004}.\footnote{These
equation are obtained at the orders $\varepsilon,\varepsilon^3$
and $\varepsilon^5$, respectively.}
    \bea
    [\rH_0,Q_1]&=&-2\rH_1,
    \label{m1}\\
    \left[\rH_0,Q_3\right]&=& -\frac{1}{6}\,[{[ \rH_1,Q_1]}, Q_1],
    \label{m2}\\
    \left[\rH_0, Q_5\right]&=&-\frac{1}{6}\,\left([[\rH_1,Q_1],Q_3]+
    [[\rH_1,Q_3],Q_1]\right)+\frac{1}{360}\,
    [[[[\rH_1,Q_1],Q_1],Q_1],Q_1].
    \label{m3}
    \eea
The higher order terms in $\varepsilon$ similarly yield operator
equations for $Q_{2i+1}$ with $i\geq 3$. As noted in
\cite{bender-prd-2004}, one can iteratively solve these equations
to obtain $Q_{2i+1}$.

A variation of the approach of \cite{bender-prd-2004} is to
substitute the ansatz\footnote{Here $\{\cdot,\cdot\}$ stands for
the anticommutator, $\{A,B\}=AB+BA$.}
    \be
    Q_{2i+1}=\sum_{j,k=0}^{i+1} c_{ijk}~\{x^{2j},p^{2k+1}\}
    \label{ansatz}
    \ee
in the operator equations for $Q_{2i+1}$ and to solve for the
coefficients $c_{ijk}$. In this way we have found the following
solutions for (\ref{m1}) and (\ref{m2}), respectively.
    \bea
    Q_1&=&
    -\frac{1}{\cum^4}\left[\frac{4}{3}\,\rp^3+\cum^2\,
    \{\rx^2,\rp\}\right]=-\frac{1}{\cum^4}\left(
    \frac{4}{3}\,\rp^3+2\cum^2\,\rx\,\rp\,\rx\right),
    \label{Q1}\\
    Q_3&=&\frac{4}{\cum^{10}}\left[
    \frac{32}{15}\,\rp^5+\frac{5}{3}\,\cum^2\{\rx^2,\rp^3\}
    +\cum^4\{\rx^4,\rp\}+2\cum^2\,\rp\right]\nn\\
     &=&\frac{128}{15\cum^{10}}\,\rp^5+\frac{40}{3\cum^8}\,
     \,\rx\,\rp^3\rx+\frac{8}{\cum^6}\, \rx^2\rp\,\rx^2-
     \frac{32}{\cum^8}\,\rp.
    \label{Q3}
    \eea
These confirm the results of \cite{bender-prd-2004} except for the
coefficient of the last term in (\ref{Q3}). We have checked the
validity of (\ref{Q3}) by inserting this relation in (\ref{m2})
and affecting both sides of the resulting equation on the function
$f_1(\rx)=\rx$. Using the fact that in the $\rx$-representation
$\rp=-i\frac{d}{d\rx}$, we could easily perform the necessary
calculations (without having to use any commutation relations) and
checked the validity (\ref{Q3}).

In fact, we can obtain the coefficients $c_{ijk}$ using this
method. In order to do this we can substitute (\ref{ansatz}) in
the operator equations for $Q_{2i+1}$, (rather than trying to use
the complicated commutation relations for powers of $\rx$ and
$\rp$) affect both sides of these equations on $f_n(\rx)=\rx^n$,
and demand that they are equal for all $n=0,1,2,3,\cdots.$

\section{The Equivalent Hermitian Hamiltonian}

Having obtained $Q$, we can easily calculate the Hermitian
Hamiltonian
    \be
    \rh=\rho\, \rH\,\rho^{-1}
    \label{rh=}
    \ee
associated with the dimensionless Hamiltonian $\rH$. Using
(\ref{rho=}), (\ref{bch}), and ({\ref{rh=}), we have
    \be
    \rh=\rH+\frac{1}{2}[\rH,Q]+\frac{1}{2!\,2^2}[[\rH,Q],Q]+
    \frac{1}{3!\,2^3}[[[\rH,Q],Q],Q]+\cdots.
    \label{bch-h}
    \ee
Now, in view of (\ref{rH}) and (\ref{Q=}), it is very easy to
identify the perturbative expansion of $\rh$, i.e., find
$\varepsilon$-independent operators $\rh^{(j)}$ such that
    \be
    \rh=\sum_{i=0}^\infty \rh^{(j)}\,\varepsilon^j.
    \label{pert-h}
    \ee
This yields
    \bea
    \rh^{(0)}&=&\rH_0,
    \quad\quad\quad\quad\rh^{(1)}=\rH_1+\frac{1}{2}\,[\rH_0,Q_1],
    \label{h01}\\
    \rh^{(2)}&=&\frac{1}{2}\,[\rH_1,Q_1]+
    \frac{1}{8}\,[[\rH_0,Q_1],Q_1],
    \label{h2}\\
    \rh^{(3)}&=&\frac{1}{2}\,[\rH_0,Q_3]+
    \frac{1}{8}\,[[\rH_1,Q_1],Q_1]+
    \frac{1}{48}\,[[[\rH_0,Q_1],Q_1],Q_1],
    \label{h3}\\
    \rh^{(4)}&=&\frac{1}{4}\,[\rH_1,Q_3]-
    \frac{1}{192}\,[[[\rH_1,Q_1],Q_1],Q_1],
    \label{h4}\\
    \rh^{(5)}&=&\frac{1}{2}\,[\rH_0,Q_5]+
    \frac{1}{12}\,([[\rH_1,Q_1],Q_3]+[[\rH_1,Q_3],Q_1])+
    \frac{1}{120}\,[[[\rH_0,Q_3],Q_1],Q_1].
    \label{h5}
    \eea
In view of the fact that $Q_1,Q_3,Q_5$ are Hermitian while $\rH_1$
is anti-Hermitian, it is not difficult to see that the terms
contributing to $\rh^{(j)}$ with even $j$ are Hermitian while
those contributing to $\rh^{(j)}$ with odd $j$ are anti-Hermitian.
The fact that $\rh$ is a Hermitian operator then suggests that the
$\rh^{(j)}$ with odd $j$ must vanish. There is another argument
supporting this expectation namely that because $\rH_1$ is a cubic
potential, the perturbation series for the ground state energy of
$\rH$ and consequently (the isospectral operator) $\rh$ must only
include even powers of the perturbation parameter $\varepsilon$,
\cite{bender-dunne}.

Using (\ref{m1}), (\ref{m2}), (\ref{m3}), we can easily show that
indeed $\rh^{(1)}$, $\rh^{(3)}$, and $\rh^{(5)}$ vanish
identically. This may be viewed as a consistency check of our
calculations. The perturbative expansion of $\rh$ valid up to and
including terms of order $\varepsilon^5$ is, therefore, given by
    \bea
    &&\rh=\rH_0+\rh^{(2)}\varepsilon^2+\rh^{(4)}\varepsilon^4
    +{\cal O}(\varepsilon^6),
    \label{rh-pert=}\\
    &&\rh^{(2)}=\frac{1}{4}\,[\rH_1,Q_1],
    \label{rh-pert-2}\\
    &&\rh^{(4)}=\frac{1}{4}\,[\rH_1,Q_3]-\frac{1}{192}\,
    [[[\rH_1,Q_1],Q_1],Q_1],
    \label{rh-pert-4}
    \eea
where we have made use of (\ref{m1}). Next, we use (\ref{Q1}) and
(\ref{Q3}) to obtain the explicit form of $\rh^{(2)}$ and
$\rh^{(4)}$. After a lengthy calculation, we find
    \bea
    &&[H_1,Q_1]=\frac{6}{\cum^4}\,\left(\{\rx^2,\rp^2\}+
    \cum^2\rx^4+\frac{2}{3}\right),
    \label{id1}\\
    &&[H_1,Q_3]=-\frac{4}{\cum^{10}}\left(16\{\rx^2,\rp^4\}
    +15\cum^2\{\rx^4,\rp^2\}+64\,\rp^2+6\cum^4\,\rx^6+
    76\cum^2\,\rx^2\right),
    \label{id2}\\
    &&[[[H_1,Q_1],Q_1],Q_1]=-\frac{48}{\cum^{12}}\left(
    8\rp^6-8\cum^2\{\rx^2,\rp^4\}+9\cum^4\{\rx^4,\rp^2\}-
    68\cum^2\rp^2+10\cum^6\rx^6+28\cum^4\rx^2\right).\nn\\
    &&
    \label{id3}
    \eea
Therefore, in view of (\ref{rh-pert-2}) and (\ref{rh-pert-4}),
    \bea
    \rh^{(2)}&=&\frac{3}{2\cum^4}
    \left(\{\rx^2,\rp^2\}+\cum^2\rx^4+\frac{2}{3}\right),
    \label{rh-2=}\\
    \rh^{(4)}&=&\frac{2}{\cum^{12}}
    \left(\rp^6-9\cum^2\{\rx^2,\rp^4\}-\frac{51}{8}\cum^4
    \{\rx^4,\rp^2\}-\frac{81}{2}\cum^2\rp^2-\frac{7}{4}\cum^6\rx^6
    -\frac{69}{2}\cum^4\rx^2
    \right).
    \label{rh-4=}
    \eea

A simple application of (\ref{rh-pert=}) is in the calculation of
the energy eigenvalues $\rE_n$ of the Hamiltonian $\rH$. If we
denote by $|n\kt$ the normalized eigenvectors of the harmonic
oscillator Hamiltonian $\rH_0$, then we can easily calculate
$\rE_n$ up to and including terms of order $\varepsilon^3$. This
is done using the first order Rayleigh-Schr\"odinger perturbation
theory which yields
    \be
    \rE_n=\cum(n+\frac{1}{2})+\br n|\rh^{(2)}|n\kt+{\cal
    O}(\varepsilon^4).
    \label{rE-pert}
    \ee
Substituting (\ref{rh-2=}) in this relation and doing the
necessary algebra, we find
    \be
    \rE_n=\cum(n+\frac{1}{2})+\frac{1}{8\cum^4}\,
    (30 n^2+30 n+11)\,\varepsilon^2+{\cal
    O}(\varepsilon^4).
    \label{rE-pert=}
    \ee
This is in complete agreement with the earlier calculations
reported in \cite{dp,bmw}.

Next, we use (\ref{new1}) - (\ref{new3}) to obtain the expression
for the unscaled Hermitian operator $h$ that is associated with
the original Hamiltonian $H$. This results in
    \bea
    h&=&\frac{p^2}{2m}+\frac{1}{2}\mu^2x^2+
    \frac{3}{2\mu^4}\left(\frac{1}{m}\,\{x^2,p^2\}+\mu^2x^4
    +\frac{2\hbar^2}{3m}\right)
    \epsilon^2+\frac{2}{\mu^{12}}
    \left(\frac{p^6}{m^3}-\frac{9\mu^2}{m^2}\{x^2,p^4\}\right.
    \nn\\
    &&
    \left. -\frac{51\mu^4}{8m}
    \{x^4,p^2\}-\frac{7\mu^6}{4}\,x^6
    -\frac{81\hbar^2\mu^2}{2m^2}\,p^2-
    \frac{69\hbar^2\mu^4}{2m}\,x^2\right)\epsilon^4+
    {\cal O}(\epsilon^6)
    \label{h-pert=4}\\
    &=&\frac{p^2}{2m}+\frac{1}{2}\mu^2x^2+
    \frac{1}{m\mu^4}\left(\{x^2,p^2\}+p\,x^2p+
    \frac{3m\mu^2}{2}\,x^4\right)
    \epsilon^2+\frac{2}{\mu^{12}}
    \left(\frac{p^6}{m^3}-\frac{63\mu^2}{16m^2}\{x^2,p^4\}\right.
    \nn\\
    &&\left.
    -\frac{81\mu^2}{8m^2}\,p^2x^2p^2-\frac{33\mu^4}{16m}
    \{x^4,p^2\}-\frac{69\mu^4}{8m}\,x^2p^2x^2
    -\frac{7\mu^6}{4}\,x^6
    \right)\epsilon^4+
    {\cal O}(\epsilon^6),
    \label{h-pert=4b}
    \eea
where we have used the identities
    \[p\,x^2p-\frac{1}{2}\{x^2,p^2\}=\hbar^2,~~~~
    x^2p^2x^2-\frac{1}{2}\{x^4,p^2\}=4\hbar^2x^2,~~~~
    p^2x^2p^2-\frac{1}{2}\{x^2,p^4\}=4\hbar^2p^2.\]

Note that if one does not truncate the perturbation expansion of
$h$, one finds that it is an infinite series in powers of $p$.
This confirms the assertion that the Hermitian Hamiltonian for a
non-Hermitian Hamiltonian with a real spectrum is in general a
nonlocal (pseudo-differential) operator, \cite{jpa-2003,p61}. A
remarkable property of the cubic anharmonic oscillator (\ref{H}),
is that the corresponding Hermitian Hamiltonian $h$ turns out to
be a local (differential) operator once one truncates its
perturbation expansion. This is not generally the case.

\section{Physical Observables}

The calculation of the physical observables $O:{\cal H}_{\rm
phys}\to{\cal H}_{\rm phys}$ mimics that of $h$. As we discussed
in Section~1, because the reference Hilbert space ${\cal H}$ for
the system is $L^2(\R)$, the observables $O$ are obtained from the
Hermitian operators $o:L^2(\R)\to L^2(\R)$ according to
(\ref{obs}). Substituting (\ref{rho=}) in this relation and using
(\ref{bch}), we have
    \be
    O=o-\frac{1}{2}[o,Q]+\frac{1}{2!\,2^2}[[o,Q],Q]
    -\frac{1}{3!\,2^3}[[[o,Q],Q],Q]\pm\cdots.
    \label{O=O}
    \ee
Moreover due to the particular $\varepsilon$-dependence of $Q$ as
given by (\ref{Q=}), we can easily determine the following
perturbation expansion for $O$.
    \be
    O=o-\frac{1}{2}\,[o,Q_1]\,\varepsilon+
    \frac{1}{8}\,[[o,Q_1],Q_1]\,\varepsilon^2-
    \frac{1}{2}\,\left([o,Q_3]+\frac{1}{24}[[[o,Q_1],Q_1],Q_1]
    \right)\,\varepsilon^3+{\cal O}(\varepsilon^4).
    \label{o-pert}
    \ee

Next, we calculate the dimensionless $\eta_+$-pseudo-Hermitian
position and momentum operators,
    \be
    \rX:=\rho^{-1}\rx\,\rho=\ell^{-1}X,~~~~~~~
    \rP:=\rho^{-1}\rp\,\rho=\ell\hbar^{-1}P.
    \label{rx-rp}
    \ee
This is done by substituting $\rx$ and $\rp$ for $o$ in
(\ref{o-pert}). Doing the necessary calculations, we obtain
    \bea
    \rX &=&\rx+\frac{2i}{\cum^4}\,\left(\rp^2+\frac{1}{2}\,
    \cum^2\rx^2\right)\varepsilon+
    \frac{1}{\cum^6}\,\left(\{\rx,\rp^2\}-\cum^2\rx^3\right)
    \varepsilon^2+{\cal O}(\varepsilon^3),
    \label{rX=}\\
    \rP &=&\rp-\frac{i}{\cum^2}\,\{\rx,\rp\}\,\varepsilon+
    \frac{1}{\cum^6}\,\left(2\rp^3-\frac{1}{2}\,\cum^2
    \{\rx^2,\rp\}\right)\varepsilon^2+{\cal O}(\varepsilon^3).
    \label{rP=}
    \eea
We can directly read the expression for the
$\eta_+$-pseudo-Hermitian position operator $X$ and momentum
operator $P$ from these equations provided that we let $\rX\to X$,
$\rP\to P/\sqrt m$, $\rx\to x$, $\rp\to p/\sqrt m$, $\cum\to\mu$,
$\varepsilon\to\epsilon$. As expected, $X$ and $P$ do not involve
$\hbar$.

Equations (\ref{rX=}) and (\ref{rP=}) show that, as operators
acting in $L^2(\R)$, $X$ and $P$ are not Hermitian. The fact that
by construction they are $\eta_+$-pseudo-Hermitian implies that as
operators acting in ${\cal H}_{\rm phys}$ they are Hermitian,
\cite{p1}. Furthermore, these operators furnish an irreducible
unitary representation of the Heisenberg-Weyl algebra,
$[X,P]=i\hbar$, on the physical Hilbert space ${\cal H}_{\rm
phys}$. They form an irreducible set of basic operators for the
quantum system, i.e., other observables may be constructed as
power series in $X$ and $P$. For instance, we can express the
Hamiltonian~(\ref{H}) according to
    \bea
    H&=&\frac{P^2}{2m}+\frac{1}{2}\mu^2X^2+
    \frac{1}{m\mu^4}\left(\{X^2,P^2\}+P\,X^2P+
    \frac{3m\mu^2}{2}\,X^4\right)
    \epsilon^2+\frac{2}{\mu^{12}}
    \left(\frac{P^6}{m^3}-\frac{63\mu^2}{16m^2}\{X^2,P^4\}\right.
    \nn\\
    &&\left.-\frac{81\mu^2}{8m^2}\,P^2X^2P^2
    -\frac{33\mu^4}{16m}\{X^4,P^2\}-
    \frac{69\hbar^2\mu^4}{8m}\,X^2P^2X^2
    -\frac{7\mu^6}{4}\,X^6\right)\epsilon^4+
    {\cal O}(\epsilon^6),
    \label{H-pert=4}
    \eea
where we have made use of (\ref{h=}), (\ref{X-P}), and
(\ref{h-pert=4b}). This is the manifestly Hermitian representation
of the original Hamiltonian~(\ref{H}).

Another interesting implication of Eqs.~(\ref{rX=}) and
(\ref{rP=}) is that if $\epsilon\neq 0$, the physical position
($X$) and momentum ($P$) operators do not satisfy the
transformation rules of the usual position ($x$) and momentum
$(p)$ operators under ${\cal P}$ and ${\cal T}$ separately,
    \[{\cal P}X{\cal P}\neq -X,~~~~{\cal P}P{\cal P}\neq -P,~~~~
    {\cal T}X{\cal T}\neq T,~~~~{\cal T}P{\cal T}\neq -P.\]
However, they share the same transformation rule under ${\cal
PT}$,
    \[{\cal PT}\,X\,{\cal PT}=-X,~~~~~~{\cal PT}\,P\,
    {\cal PT}= P.\]
This is consistent with the fact that unlike ${\cal P}$ and ${\cal
T}$, ${\cal PT}$ is an antilinear $\eta_+$-pseudo-unitary operator
\cite{jmp-2004}.\footnote{This can be easily checked using the
approach of \cite{jmp-2003}.} In particular, it implies that, as
an operator acting in ${\cal H}_{\rm phys}$, ${\cal PT}$ is an
antilinear unitary operator. This in turn implies, in view of
Wigner's classification of symmetries in quantum mechanics
\cite{weinberg}, that unlike ${\cal P}$ and ${\cal T}$, ${\cal
PT}$ defines a physical symmetry of the quantum system. The fact
that ${\cal P}$ does not correspond to a physical symmetry was to
be expected, for its definition is intertwined with that of $x$
which is not a physical observable for the system unless
$\epsilon=0$.

\section{The Classical Limit}

The phase space of the underlying classical Hamiltonian for the
cubic anharmonic oscillator (\ref{H}) is clearly $\R^2$. Having
calculated the Hermitian operator $h$ we can determine the
classical Hamiltonian $H_c$ for this system using (\ref{class}).
In view of (\ref{h-pert=4}), the evaluation of the limit in
(\ref{class}) is trivial. Up to and including terms of order
$\epsilon^5$, $H_c$ is given by
    \bea
    H_c&=&\frac{p_c^2}{2m}+\frac{1}{2}\mu^2x_c^2+
    \frac{3}{2\mu^4}\left(\frac{2}{m}\,x_c^2p_c^2+\mu^2x_c^4\right)
    \epsilon^2+\nn\\
    &&\frac{2}{\mu^{12}}
    \left(\frac{p_c^6}{m^3}-\frac{18\mu^2}{m^2}\,x_c^2p_c^4-
    \frac{51\mu^4}{4m}
    \,x_c^4p_c^2-\frac{7\mu^6}{4}\,x_c^6\right)\epsilon^4+
    {\cal O}(\epsilon^6).
    \label{H-class=}
    \eea

We shall first explore the consequences of neglecting the terms of
order $\epsilon^4$ and higher. Then we can express $H_c$ in the
form
    \bea
    H_c&=&\frac{p_c^2}{2M(x_c)}+\frac{\mu^2}{2}\,x_c^2+
    \frac{3\epsilon^2}{2\mu^2}\,x_c^4+{\cal O}(\epsilon^4),
    \label{H-class=2}\\
    M(x_c)&:=&\frac{m}{1+3\mu^{-4}\epsilon^2\,x_c^2}=
    m(1-3\mu^{-4}\epsilon^2\,x_c^2)+{\cal O}(\epsilon^4).
    \label{M}
    \eea
Therefore, for sufficiently small $\epsilon$, $H_c$ describes the
dynamics of a point particle with a position-dependent mass
$M(x_c)$ that interacts with a quartic anharmonic potential. This
statement provides a physical interpretation of the original
${\cal PT}$-symmetric cubic anharmonic oscillator (\ref{H}).
Obviously, this is a valid approximation as long as we can neglect
the contribution from the terms of order $\epsilon^4$ and higher,
${\cal O}(\epsilon^4)\approx 0$.

Under this assumption, $H_c$ takes non-negative values; the
classically allowed energies $E$ are non-negative. This is the
classical analog of the fact that the ${\cal PT}$-symmetric
quantum Hamiltonian (\ref{H}) has a positive spectrum. Moreover,
it is not difficult to show that the classical orbits in the phase
space for the Hamiltonian (\ref{H-class=2}) are ellipses
determined by
    \be
    \frac{p_c^2}{2m}+\left(\frac{\mu^2}{2}+\frac{3\epsilon^2E}{\mu^4}
    \right)\,x_c^2=E.
    \label{ellips}
    \ee
The coupling of the energy $E$ and the perturbation parameter
$\epsilon$ is an indication that the above approximation is valid
for low energies, i.e.,
    \be
    E\ll E_\star:=\frac{1}{6}\mu^6\epsilon^{-2}.
    \label{E-bnd}
    \ee

The inclusion of the terms of order $\epsilon^4$ distorts the
above picture. However, as long as condition (\ref{E-bnd}) holds
the classical (phase-space) orbits are closed curves.
Figure~\ref{fig1} shows the graph of such orbits.
    \begin{figure}[ht]
    \vspace{.5cm}
    \centerline{\epsffile{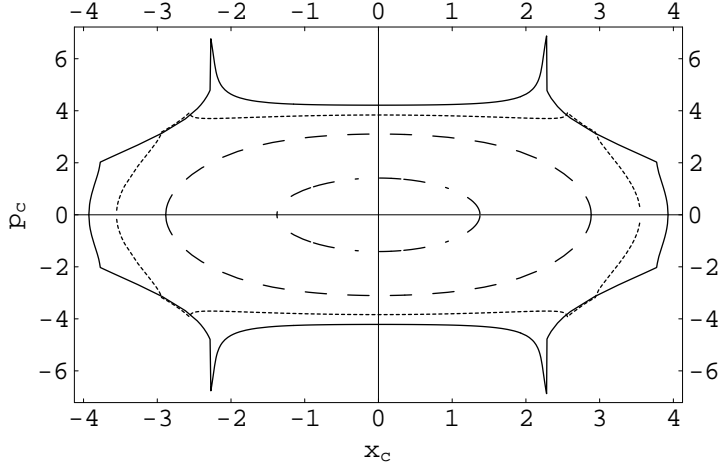}}
    \vspace{.5cm}
    \centerline{
    \parbox{14cm}{\caption{Graph of the orbits in phase space
    for the Hamiltonian (\ref{H-class=}) with ${\cal O}(\epsilon^6)$
    neglected, $m=\mu=1$, $\epsilon=0.1$, and $E=1$ (dashed-dotted
    curve),
    $E=5$ (dashed curve), $E=8$ (dotted curve),
    and $E=10$ (full curve). The horizontal and vertical axes
    are those of $x_c$ and $p_c$, respectively. Note that
    $E_\star=50/3\approx 16.7$. Hence our perturbative calculation
    of the classical orbit for $E=10$ is not as reliable as that for
    $E=1$ and $E=5$. In particular the elliptic shape of the $E=1$
    orbit is consistent with Eq.~(\ref{ellips}).}\label{fig1}}}
    \end{figure}

If we perform the $\eta_+$-pseudo-Hermitian quantization of the
classical Hamiltonian (\ref{H-class=}), namely let $x_c\to X$,
$p_c\to P$, and $\{\cdot,\cdot\}_c\to-i\hbar^{-1}[\cdot,\cdot]$,
where $\{\cdot,\cdot\}_c$ is the classical Poisson bracket, we
recover the expression~(\ref{H-pert=4}) for the original
Hamiltonian (\ref{H}), provided that we adopt the correct
factor-ordering prescription. This observation underlines the
importance of the issue of factor-ordering ambiguity in
pseudo-Hermitian and in particular ${\cal PT}$-symmetric quantum
mechanics.

\section{The Conserved Probability Density}

The expression (\ref{rho}) for invariant probability density for
the localization of the quantum system under consideration
involves the physical wave function (\ref{wf}). Given a state
vector $\psi\in{\cal H}_{\rm phys}$, the perturbation expansion
for the corresponding physical wave function is given by
    \bea
    \Psi(x)&=&\br x|e^{-Q/2}|\psi\kt=\br x|\sum_{k=0}^\infty
    \frac{(-1)^kQ^k}{2^k k!}|\psi\kt\nn\\
    &=&\psi(x)-\mbox{\small$\frac{1}{2}$}\br x|Q_1|\psi\kt\varepsilon+
    \mbox{\small$\frac{1}{8}$}\br x|Q_1^2|\psi\kt\varepsilon^2
    -\mbox{\small$\frac{1}{2}$}\left(\br x|Q_3|\psi\kt+
    \mbox{\small$\frac{1}{24}$}\,
    \br x|Q_1^3|\psi\kt\right)\varepsilon^3+
    {\cal O}(\varepsilon^4),
    \label{wf=}
    \eea
where we have used (\ref{rho=}) and (\ref{Q=}). We can obtain the
explicit form of the terms appearing on the right-hand side of
(\ref{wf=}) using (\ref{Q1}), (\ref{Q3}), (\ref{new1}) --
(\ref{new3}), and the identity $\br x|p=-i\hbar\frac{d}{dx}\br
x|$. The result may be expressed as
    \be
    \Psi(x)=(1+\epsilon\hat L_1+\epsilon^2
    \hat L_2+\epsilon^3\hat L_3)\psi(x)+{\cal O}(\epsilon^4),
    \label{wf=2}
    \ee
where
    {\small\bea
    &&\hat L_1:=-\frac{1}{2}\,\hat Q_1,~~~~
    \hat L_2:=\frac{1}{8}\,\hat Q_1^2,~~~~
    \hat L_3:=-\frac{1}{2}\,\hat Q_3-\frac{1}{48}\,\hat Q_1^3,~~~~
    \hat Q_1:=\frac{2i}{\mu^4}\left[-\frac{2\hbar^2}{3m}\,
    \frac{d^3}{dx^3}+\mu^2(x^2\frac{d}{dx}+x)\right],\nn\\
    &&\hat Q_3:=\frac{4i}{\mu^{10}}\left[-\frac{32\hbar^4}{15m^2}\,
    \frac{d^5}{dx^5}+\frac{10\hbar^2\mu^2}{3m}\left(x^2\frac{d^3}{dx^3}+
    3x\,\frac{d^2}{dx^2}\right)-2\mu^4\left(x^4\frac{d}{dx}+
    2x^3\right)+\frac{8\hbar^2\mu^2}{m}\,\frac{d}{dx}\right].\nn
    \eea}

Having obtained the general form of the physical wave function we
can calculate the invariant probability density $\varrho$
according to (\ref{rho}). Figures~3 and~4 show the plots of
$\varrho$ for $\psi(x)=e^{-x^2/2}$ and $x\,e^{-x^2/2}$.
    \begin{figure}[p]
    \centerline{\epsffile{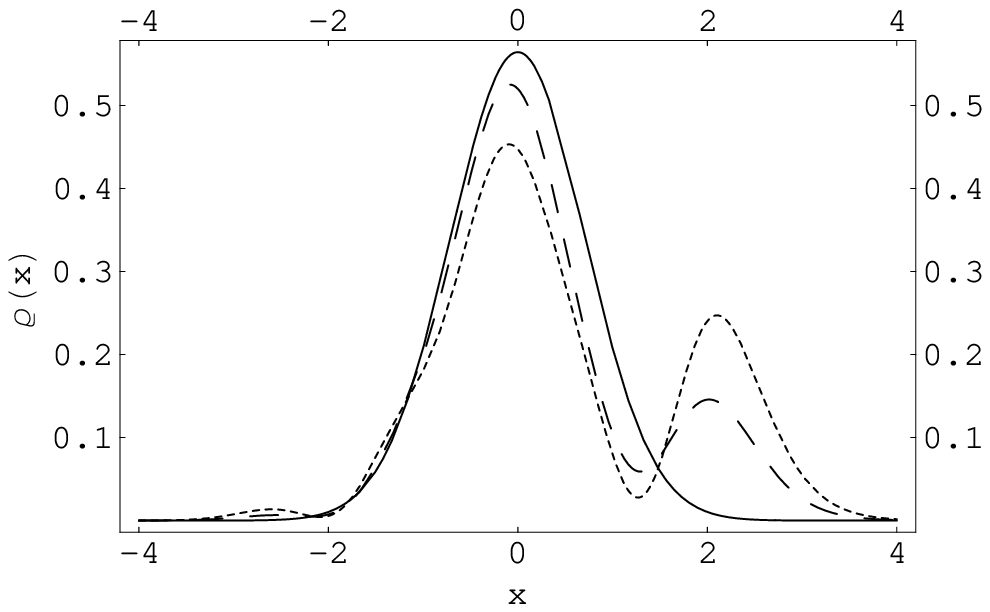}}
    \centerline{
    \parbox{12cm}{\caption{Graph of the invariant probability density
    $\varrho$ for $\psi(x)=e^{-x^2/2}$, $\hbar=m=\mu=1$ and
    $\epsilon=0$ (full curve) $\epsilon=0.2$ (dashed curve),
    and $\epsilon=0.25$ (dotted curve).}\label{fig2}}}
    \vspace{2cm}
    \centerline{\epsffile{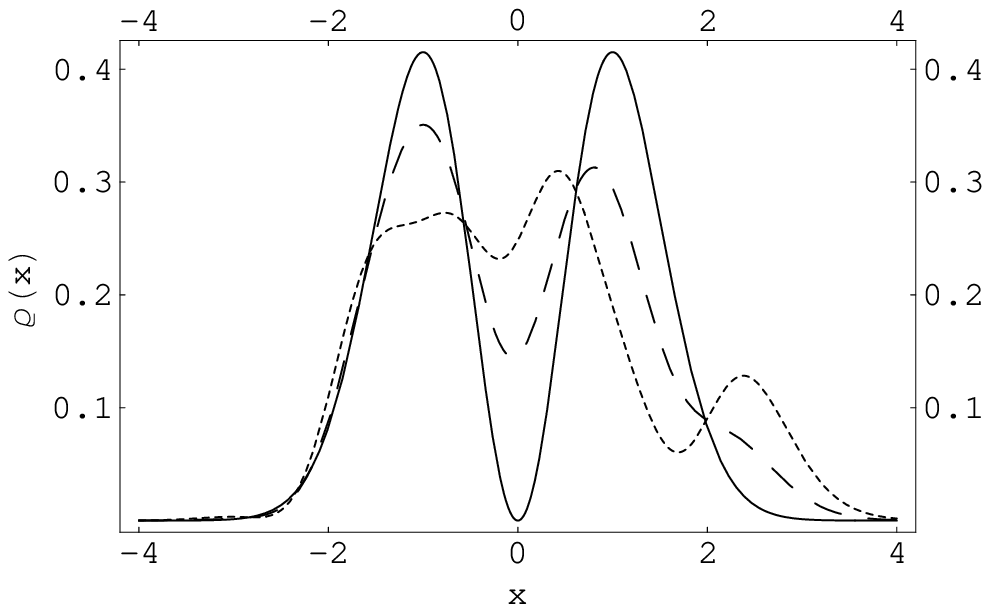}}
    \centerline{
    \parbox{12cm}{\caption{Graph of the invariant probability density
    $\varrho$ for $\psi(x)=x\,e^{-x^2/2}$, $\hbar=m=\mu=1$ and
    $\epsilon=0$ (full curve) $\epsilon=0.1$ (dashed curve),
    and $\epsilon=0.15$ (dotted curve).}\label{fig4}}}
    \end{figure}

\section{Conclusion}

We have performed a comprehensive study of the physical content of
the ${\cal PT}$-symmetric quantum system based on the
non-Hermitian cubic anharmonic oscillator~(\ref{H}). We showed how
the general ideas developed within the framework of the
pseudo-Hermitian quantum mechanics may be applied to this model.
The result is an explicit characterization of the corresponding
Hermitian Hamiltonian, physical observables, probability density,
and the underlying classical system. The only other ${\cal
PT}$-symmetric system (with an infinite-dimensional state space)
for which a similar treatment has been possible is the ${\cal
PT}$-symmetric square well \cite{p61}. An important difference
between the latter system and the anharmonic oscillator~(\ref{H})
is that the effects of non-Hermiticity of this oscillator do
survive the classical limit; non-Hermiticity is not a by-product
of the (pseudo-Hermitian) quantization.

Neglecting forth and higher order terms in our perturbative
treatment, we showed that the ${\cal PT}$-symmetric cubic
anharmonic oscillator~(\ref{H}) describes a point particle having
a position-dependent mass and interacting with a real quartic
anharmonic potential. This provides a classical justification for
the positivity of the spectrum of (\ref{H}). The same argument
applies to the cases where we should keep the terms of order up to
(and including) five.

The pseudo-Hermitian quantization of the classical Hamiltonian
defined by the appropriate metric operator together with a
particular factor-ordering prescription yields the original local
${\cal PT}$-symmetric Hamiltonian while the usual canonical
quantization of the same classical Hamiltonian with the
appropriate factor-ordering prescription leads to the
corresponding equivalent nonlocal Hermitian Hamiltonian.

The approach pursued in this paper may be applied to other ${\cal
PT}$-symmetric and non-${\cal PT}$-symmetric non-Hermitian
Hamiltonians with a real spectrum. In general, however, the
nonlocality of the corresponding equivalent Hermitian Hamiltonian
may manifest itself at each order of the perturbation theory. This
has already been the case for the ${\cal PT}$-symmetric square
well studied in \cite{p61}. In view of the results of
\cite{bender-prd-2004}, the same is the case for the ${\cal
PT}$-symmetric cubic potential, i.e., (\ref{H}) with $\mu=0$. An
interesting subject of future study is to extend the approach
pursued here to the field theoretical analog of (\ref{H}). Such a
study should reveal the structure of the underlying classical
field theory. \vspace{.3cm}

\noindent {\bf Note:} After the online announcement of the
preprint of this article (quant-ph/0411137), Hugh Jones sent me
his preprint: quant-ph/0411171 in part of which he also studies
the ${\cal PT}$-symmetric cubic anharmonic oscillator.

\section*{Acknowledgment}

I wish to thank Bijan Bagchi and Hugh Jones for their comments
that helped me correct the minor errors in an earlier version of
the paper.

\section*{Erratum}

There is a factor of 2 error in Eq.~(61) which was noticed after
the publication of the paper. Correcting this error leads to minor
changes in Eqs.~(62) and (63). The corrected equations are
    \[
    \begin{array}{ccc}
    M(x_c):=\frac{m}{1+6\mu^{-4}\epsilon^2x_c^2}=
    m(1-6\mu^{-4}\epsilon^2x_c^2)+{\cal O}(\epsilon^4),
    &~~~~~~~~~~~~~~~~~~~~~~~~~&(61)\\
    \frac{p_c^2}{2m}+\left(\frac{\mu^2}{2}+
    \frac{6\epsilon^2E}{\mu^4}
    \right)\,x_c^2-\frac{3\epsilon^2}{2\mu^2}\,x_c^4=E,
    &~~~~~~~~~~~~~~~~~~~~~~~~~&(62)\\
    E\ll E_\star:=\frac{1}{12}\mu^6\epsilon^{-2}.
    &~~~~~~~~~~~~~~~~~~~~~~~~~&(63)\end{array}\]
Eq.~(62) shows that the distortion of the elliptic shape of the
phase space orbits of the unperturbed (harmonic oscillator)
potential occurs at order $\epsilon^2$ of the perturbation theory.
This distortion is more pronounced for larger values of $E$ as
shown in Figure~1. Note that this figure uses Eq.~(59) which is
free from the above-mentioned numerical error. \vspace{.5cm}

I wish to thank Christiane Quesne for informing me of the above
error.



\ed

\begin{figure}
\centerline{\epsffile{g.eps}} \caption{graph}
\end{figure}